\documentclass[sigconf]{acmart}
\AtBeginDocument{%
  }

\renewcommand\footnotetextcopyrightpermission[1]{}

\newif\ifreview
\reviewfalse

\ifreview
  \usepackage[switch]{lineno}
\fi

\usepackage{balance}
\usepackage{tcolorbox}
\usepackage{iftex}
\usepackage[english]{babel}
\usepackage{listofitems}
\usepackage{xspace}

\makeatletter
\g@addto@macro{\UrlBreaks}{\UrlOrds}
\makeatother

\usepackage{tikz}
\tikzstyle{state}=[draw=black, thin, circle, minimum size=10pt, inner sep=0pt, fill=none]
\tikzstyle{none}=[fill=none, draw=none]
\tikzstyle{pill}=[draw=gray, drop shadow,very thin,dotted,rounded corners=3pt,fill=lightgray!20,inner sep=3pt]
\tikzstyle{alertpill}=[draw=red,thick,dashed,rounded corners=5pt,fill=red!20,inner sep=10pt]
\tikzstyle{left_item}=[draw=black, drop shadow, outer sep=0mm, fill=lightgray!20, inner ysep = 2mm,  minimum width = 5cm]
\tikzstyle{source}=[fill=NavyBlue!10, drop shadow, inner sep = 4pt, minimum width = 4.5cm]

\tikzstyle{std-egde}=[draw=black, ->]
\tikzstyle{double-edge}=[->, double]
\tikzstyle{dotted-line}=[-, dotted]

\usetikzlibrary{calc}
\usetikzlibrary{fit,backgrounds}

\pgfdeclarelayer{nodelayer}
\pgfdeclarelayer{edgelayer}
\pgfdeclarelayer{background}
\pgfsetlayers{background,nodelayer,main,edgelayer}
\usetikzlibrary{shapes.symbols, arrows.meta, patterns, patterns.meta, automata, positioning, decorations, decorations.pathreplacing, decorations.pathmorphing, fit, calc, shadows}
\usetikzlibrary{snakes,arrows,shapes}

\usepackage{amsmath}
\usepackage{amsthm}
\usepackage{fontawesome}

\usepackage{float}
\usepackage{graphicx}
\usepackage{diagbox}
\usepackage[dvipsnames,svgnames,x11names,table]{xcolor}
\usepackage{listings}
\definecolor{codebg}{gray}{0.96}
\definecolor{codecomment}{gray}{0.45}
\definecolor{codekw}{rgb}{0.13,0.13,0.55}
\usepackage[autostyle=true]{csquotes}

\usepackage[normalem]{ulem}

\usepackage{threeparttable}
\usepackage{booktabs}
\usepackage{paralist}
\usepackage{amsmath}

\usepackage[]{natbib}

\usepackage{algorithm}
\usepackage{algorithmicx}
\usepackage{algpseudocode}

\usepackage{multirow}
\usepackage{dcolumn}

\usepackage{pdfcomment}

\makeatletter
\@ifundefined{missingfigure}{}{}
\@ifundefined{textcomment}{}{}
\@ifundefined{sidecomment}{}{}
\@ifundefined{todo}{}{}
\@ifundefined{TODO}{}{}
\@ifundefined{todofix}{}{}
\@ifundefined{change}{}{}
\@ifundefined{done}{}{}
\makeatother

\providecommand{\unislop}{\textsc{unislop}\xspace}

\usepackage[group-minimum-digits=4,per-mode=fraction]{siunitx}

\usepackage{hyperref}
\hypersetup{hidelinks, colorlinks=true, raiselinks=true, allcolors=black, pdfstartview=Fit, breaklinks=true, hypertexnames=false}

\usepackage[all]{hypcap}
\usepackage[caption=false,font=footnotesize]{subfig}
\usepackage[capitalise,nameinlink,noabbrev]{cleveref}

\crefformat{section}{\S#2#1#3}
\crefformat{sections}{\S#2#1#3}
\crefformat{subsection}{#2#1#3}
\crefmultiformat{subsection}{#2#1#3}{,~#2#1#3}{,~#2#1#3}{}
\crefformat{subsubsection}{\S#2#1#3}
\crefrangeformat{section}{\S\S#3#1#4 to~#5#2#6}

\crefname{figure}{Fig.}{Fig.}
\crefname{listing}{List.}{List.}
\Crefname{listing}{List.}{List.}
\crefname{lstlisting}{List.}{List.}
\Crefname{lstlisting}{List.}{List.}

\definecolor{amber}{rgb}{1.0, 0.75, 0.0}
\newcounter{ReviewerID}
\readlist*\annotationcolors{Blue, Red, Orange, Green, Purple}
\newcommand{\newreviewer}[2]{%
        \ifnum \theReviewerID=\annotationcolorslen
        \setcounter{ReviewerID}{0}
        \fi
        \stepcounter{ReviewerID}%
        \expandafter\edef\csname bootstrap#1\endcsname{%
                \expandafter\def\csname #1\endcsname####1{%
                        \ifreview%
                         {\noexpand\color{\annotationcolors[\theReviewerID]} {\noexpand\bf{\noexpand\fbox{#2}} {\noexpand\it ####1} }}
                        \else%
                         {}%
                        \fi%
                }%
        }%
        \csname bootstrap#1\endcsname%
}

\newreviewer{ed}{ed}
\newreviewer{mb}{mb}
\newreviewer{mat}{mat}
\newreviewer{mao}{mao}

\usepackage{enumitem}

\begin{document}

\begin{abstract}
Baseband processors are always reachable over the radio.
Their most security-critical logic runs deep inside protocol state machines: the control-plane handlers that gate registration, authentication, and session setup.
Analyzing this complex logic systematically requires introspecting the running firmware, which makes re-hosting the baseband necessary.
Existing re-hosting work approximates the execution environment and under-approximates the SoC complexity of the baseband processor together with its surrounding components, bringing this state practically out of reach.
We instead model each surrounding component, co-processors, SIM, application processor, from what a real device does, and execute them in lockstep with the baseband on one shared clock.
Our approach achieves high fidelity at component interfaces.

We call this method \unislop and demonstrate it on the UNISOC UDX710, a platform in an estimated 10--15\% of cellular modems and in automotive systems, not systematically analyzed before.
Starting from a Quectel RM500U-CNV module, we gain code execution, defeat its firmware-integrity check, instrument the baseband, and recover its peripheral environment from the running device.
The resulting re-hosted environment reaches the same control-plane states as the real device, establishes a full PDU session, and carries real IP traffic on both ingress and egress.
The recovered components are shared across UNISOC's baseband lineup, so with additional reverse-engineering effort the same design extends to further targets.
\end{abstract}

\title{\enquote{Operator, can you hear me?}\\A Faithful Line into the UNISOC Baseband}

\author{Eduard Vlad, Philipp Mao, Marcel Busch, Haitham Hassanieh, Mathias Payer}
\affiliation{%
  \institution{EPFL, Switzerland}
  \country{}
}

\maketitle

\ifreview%
  \linenumbers
  \pagestyle{plain}
\else%
 {}%
  \pagestyle{plain}
\fi%

\makeatletter
\newcommand{\inlinesubsection}{%
  \@ifstar{\inlinesubsection@disabled}{\inlinesubsection@enabled}%
}

\newcommand{\inlinesubsection@disabled}[1]{%
  \noindent\textbf{#1}%
}

\newcommand{\inlinesubsection@enabled}{%
  \@ifnextchar[{\inlinesubsection@opt}{\inlinesubsection@opt[]}%
}

\def\inlinesubsection@opt[#1]#2#3{%
  \refstepcounter{subsection}%
  \noindent\textbf{#1\arabic{subsection}\ #2}%
  \label{#3}%
}
\makeatother
\newcommand\designref[1]{D\cref{#1}}

\providecommand{\unislop}{\textsc{unislop}\xspace}
\providecommand{\baseslop}{\textsc{BaseSLOP}\xspace}
\section{Introduction}
\label{sec:introduction}

Cellular basebands are always-on, radio-reachable computers embedded in phones, vehicles, routers, and industrial IoT~\cite{quectel_rm500u_series,samsung_baseband,qualcomm_baseband,mediatek_baseband}.
They expose complex functionality in the form of stateful protocol handling.
The handlers that matter, such as 5G RRC and NAS, EAP-AKA$'$, and SIP/IMS, run only after a sequence of messages has driven the stack into an accepting state.
Reaching and exercising these deep state machines on real firmware is largely unexplored, because it requires running the firmware itself, faithfully, all the way into those states.
Reaching precise, high-fidelity emulation of complex firmware remains an unsolved problem.\par

\textbf{Prior Art.}
Prior work reconstructs the baseband by approximating its functionality and its execution context. As a consequence, only a subset of its functionality can be exercised.
BaseBridge~\cite{klischies2025basebridge} snapshots execution through vendor debug features, but abstracts critical system-on-chip (SoC) details and thus cannot follow the message sequences that induce critical state transitions.
FirmWire~\cite{hernandez2022firmwire} disables the RTOS tasks with hardware dependencies, so large parts of the baseband cannot be executed.
Loris~\cite{ranjbar2025stateful} isolates a single control-plane task with concolic execution and abstracts everything that happens outside it.
Under-approximating the context drops functionality the firmware depends on, while over-approximating it does not scale and forces a narrow scope, so in both cases the functionality of interest is missed.
This happens because a baseband is not a single processor: it interacts continuously with L2 co-processors and DSPs, clocks, cryptographic engines, the SIM, and the application processor.
Reconstructing the firmware alone, without the state and behavior of these components, inevitably forces either under- or over-approximation to reach the relevant protocol surfaces.
For example, the application processor \textit{drives the baseband} to attach to a base station.
Trying to exercise protocol surfaces without an application processor (without connecting to a base station) cannot be done faithfully.\par

\textbf{Insight.}
We reconstruct baseband functionality and state handling by controlling SoC components, \textit{peers}, concretely instead of approximating them.
We drive these SoC peers and the time progression from the same clock cycles, which determines execution and lets us present each component with the behavior the firmware expects, when it expects it.
We do not only replicate the baseband to run the firmware, we build the surrounding state that the firmware acts on.
To know that the reconstruction is correct, we need a notion of faithful emulation.\par

\textbf{Challenges.}
Faithful emulation depends on three linked properties.
\textit{First}, the components must be present with correct interfaces; an absent one makes the firmware assert, and an approximated one diverges silently.
\textit{Second}, they must be exercised in the right order and timing to build the state that gates each handler, which requires the first property.
\textit{Third}, execution and protocol state must be observable, otherwise the first two cannot be ensured.
All three must hold together, and determinism is what makes them hold the same way on every run.\par

We assess these properties through fidelity, our notion of correctness.
An emulation is faithful when its interaction with the surrounding components matches the real device at the inter-SoC peripheral interfaces.
We therefore measure fidelity against real silicon at these interfaces instead of assuming it, and use it to decide whether the three properties are fulfilled.\par

\textbf{Solution and Instantiation.}
We address this with \unislop, a five-step method for gaining full and faithful execution of a UNISOC baseband: obtain code execution, run arbitrary firmware, instrument it for tracing and introspection, use that access to recover the surrounding components and their behavior on real silicon, and reconstruct them as a deterministic emulation.
We target UNISOC because it is important and undocumented: it accounts for an estimated 10--15\% of cellular modems~\cite{sni5gect}, reaches automotive head units~\cite{kaspersky_unisoc_car}, and has no published systematic analysis.
The platform is closed and protects firmware loading with a software integrity check, so obtaining code execution and loading custom firmware past this check is itself a contribution and a prerequisite for the rest.
We apply \unislop to the UNISOC UDX710 baseband and demonstrate the first emulated baseband that can exercise the entire 5G handshake and establish a full IPv4 PDU session with real ingress and egress data.\par

\textbf{Contributions.}
We present \unislop, the first systematic security analysis of a UNISOC baseband.
Our contributions are threefold: (1) we unlock and instrument a previously undocumented UNISOC baseband, including a bypass of Quectel's firmware-integrity protection; (2) we introduce peer- and interconnect-driven re-hosting, a design principle for deterministic baseband emulation with an operational definition of fidelity at the interconnect boundary; and (3) we validate \unislop on the UDX710, showing it reaches the same control-plane state transitions as the real device and carries genuine control- and data-plane traffic through a full PDU session.\par

\section{Background}
\label{sec:background}

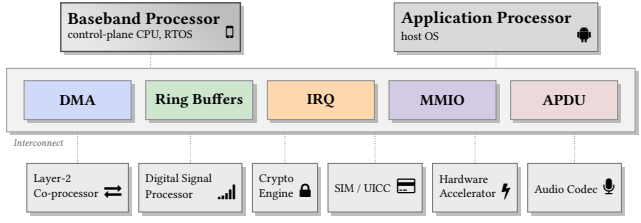
\begin{figure}
	\centering
	\resizebox{\linewidth}{!}{
		\begin{tikzpicture}[
  font=\footnotesize,
  comp/.style={draw=black!55, fill=black!8, drop shadow, inner sep=4pt, minimum height=1cm},
  cpu/.style={comp, fill=black!15},
  mech/.style={draw=black!45, drop shadow, minimum width=2.2cm, minimum height=0.8cm,
               align=center, font=\bfseries},
]
\newcommand{\cell}[2]{\begin{tabular}{@{}l@{\ \ }c@{}}#1 & \large#2\end{tabular}}

% masters
\node[cpu, shade, top color=black!26, bottom color=black!9, shading angle=-40, draw=black!70, thick] (bb) at (-3.5,1.5) {\cell{\shortstack[l]{\large\textbf{Baseband Processor}\\{control-plane CPU, RTOS}}}{\faMobile}};
\node[cpu] (ap) at ( 3.6,1.5) {\cell{\shortstack[l]{\textbf{\large Application Processor}\\{host OS}}}{\faAndroid}};

% interconnect: colored mechanism boxes
\node[mech, fill=RoyalBlue!25]   (dma) at (-5.0,0) {DMA};
\node[mech, fill=ForestGreen!22] (rb)  at (-2.5,0) {Ring Buffers};
\node[mech, fill=BurntOrange!32] (irq) at ( 0.0,0) {IRQ};
\node[mech, fill=RoyalPurple!20] (mmio)at ( 2.5,0) {MMIO};
\node[mech, fill=Maroon!15]      (apdu)at ( 5.0,0) {APDU};

% peers (two rows of three), spelled out, icon right
\node[comp] (l2)  at (-5.0,-1.8) {\cell{\shortstack[l]{Layer-2\\Co-processor}}{\faExchange}};
\node[comp] (dsp) at ($(l2.east) + (0.2,0)$) [anchor=west] {\cell{\shortstack[l]{Digital Signal\\Processor}}{\faSignal}};
\node[comp] (ce)  at ($(dsp.east) + (0.2,0)$) [anchor=west] {\cell{\shortstack[l]{Crypto\\Engine}}{\faLock}};
\node[comp] (sim) at ($(ce.east) + (0.2,0)$) [anchor=west] {\cell{\shortstack[l]{SIM / UICC}}{\faCreditCard}};
\node[comp] (acc) at ($(sim.east) + (0.2,0)$) [anchor=west] {\cell{\shortstack[l]{Hardware\\Accelerator}}{\faBolt}};
\node[comp] (aud) at ($(acc.east) + (0.2,0)$) [anchor=west] {\cell{\shortstack[l]{Audio Codec}}{\faMicrophone}};

% the interconnect band behind the mechanism boxes
\begin{pgfonlayer}{background}
  \node[draw=black!40, fill=black!5, drop shadow, inner xsep=10pt, inner ysep=7pt,
        fit=(dma)(apdu)] (bus) {};
\end{pgfonlayer}
\node[anchor=north west, font=\scriptsize\itshape, black!60]
     at ([xshift=1pt,yshift=-1pt]bus.south west) {Interconnect};

% dotted connectors from every component to the interconnect
\begin{scope}[every path/.style={draw=black!55, densely dotted, thin}]
  \draw (bb.south) -- (bb.south |- bus.north);
  \draw (ap.south) -- (ap.south |- bus.north);
  \foreach \n in {l2,dsp,ce,sim,acc,aud}{ \draw (\n.north) -- (\n.north |- bus.south); }
\end{scope}
\end{tikzpicture}
	}
	\caption{Baseband SoC. The baseband and application processors, and the co-processors and peripherals, all communicate over dedicated interconnects (DMA, ring buffers, IRQ, MMIO, APDU).
    We model the peers that gate control-plane state.}
	\label{fig:soc-arch}
\end{figure}

\textbf{Baseband Architecture.}
A baseband processor implements cellular communication and, together with the Subscriber Identity Module (SIM), forms the User Equipment (UE).
\emph{Functionally}, it runs the cellular protocol stacks in software on a real-time operating system (RTOS), with tasks for Radio Resource Control (RRC), Non-Access Stratum (NAS), and the lower layers, communicating through queues and shared memory under tight timing constraints.
\emph{Structurally}, it is not a single processor but one component of a System-on-Chip (SoC), shown in \cref{fig:soc-arch}.
An application processor (AP) drives the baseband, for instance to bring up a data or voice session.
Co-processors and peripherals perform the work the baseband offloads: a Layer-2 co-processor and a digital signal processor (DSP) for physical-layer and MAC processing, a cryptographic engine for ciphering, integrity, and key derivation, hardware accelerators for the data path, the SIM for credentials, and an audio codec for voice.
Every component communicates to the baseband over a dedicated interconnect: direct memory access (DMA), ring buffers, mailboxes with interrupts (IRQ), memory-mapped I/O (MMIO), and, for the SIM, APDU exchanges.
Some of these components can carry attacker-controlled input, for example the AP, the SIM, and the Layer-2 co-processor.
Re-hosting a baseband is not as simple as emulating its processor. The firmware's handling of each message depends on state built through timed interactions with exactly these co-processors and peers, so stubbing or approximating them invalidates that state and, with it, correct message handling.

\begin{figure}
	\centering
	\small
	\resizebox{\linewidth}{!}{
		\input{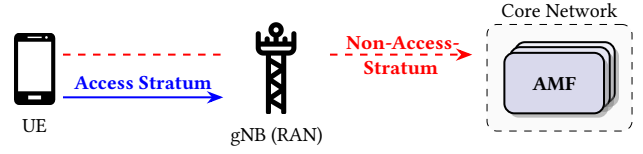}
	}
	\caption{Simplified 5G Network Architecture.}
	\label{fig:network_architecture}
\end{figure}

\textbf{Cellular Networks and Protocols.}
Beyond the UE, a cellular network comprises the Radio Access Network (RAN, e.g., a 5G gNB) that terminates the air interface, and the core network that provides authentication, mobility, session management, and data connectivity (\cref{fig:network_architecture}).
Control-plane signaling uses RRC between the UE and the RAN and NAS between the UE and the core, which the RAN relays transparently; downlink and uplink denote network-to-UE and UE-to-network messages.
A typical 5G attach proceeds from RRC connection setup through NAS registration, mutual authentication, and security-mode configuration, to data-session establishment.
Along the way, security contexts, identifiers, sequence numbers, and counters accumulate as state that gates later handling, which is exactly the state a faithful emulation must reconstruct.

\providecommand{\unislop}{\textsc{unislop}\xspace}
\providecommand{\baseslop}{\textsc{BaseSLOP}\xspace}
\begingroup
\setlist[itemize]{leftmargin=1.4em,itemsep=1pt,topsep=2pt,parsep=0pt}

\section{The UNISOC UDX710 Baseband}
\label{sec:target}

The UNISOC UDX710 is a 5G baseband platform which extends well beyond smartphones.
It ships in modem modules such as the Quectel RM500U-CNV, which integrators embed in industrial routers, home gateways, digital signage, and a range of other IoT devices~\cite{quectel_rm500u_series,kaspersky_unisoc_car}.
A vulnerability in this baseband reaches a broad ecosystem of connected devices, not only phones.
Yet UNISOC is undocumented, and to our knowledge no prior work has published a systematic analysis of a UNISOC baseband, so gaining access to and instrumenting it is itself a prerequisite we must establish.

We study the RM500U-CNV as a standalone module rather than as part of an end-user device.
It exposes USB and PCIe host interfaces; we use USB through a dedicated adapter board, which carries AT commands, ADB access once enabled, and the logging and introspection we rely on later.
For over-the-air validation we attach the modem to a controlled 5G testbed: a host PC runs the base station and core network while an SDR provides the radio link (\cref{fig:rm500u-setup}), letting us drive real network procedures while observing the modem over USB.

\textbf{Outline.}
In the following, we work through a four-step chain that makes the baseband testable and analyzable.
First, we unlock the baseband (\cref{sec:foothold}), reverse-engineering the platform and bypassing firmware verification to run modified images. 
We instrument the baseband for tracing and introspection (\cref{sec:observing}); we reconstruct the peripheral environment the firmware expects (\cref{sec:periph}); and finally we re-host the baseband as a deterministic emulation (\cref{sec:rehost}).

\begin{figure}

\usetikzlibrary{positioning,arrows.meta,decorations.pathmorphing}

\resizebox{\linewidth}{!}{
\begin{tikzpicture}[
    >=Stealth,
    node distance=1.8cm and 2.5cm,
    host/.style={
        draw,
        rectangle,
        rounded corners,
        minimum width=2.8cm,
        minimum height=1.2cm,
        align=center
    }
]

\node[] (0,0) {\includegraphics[width=\linewidth]{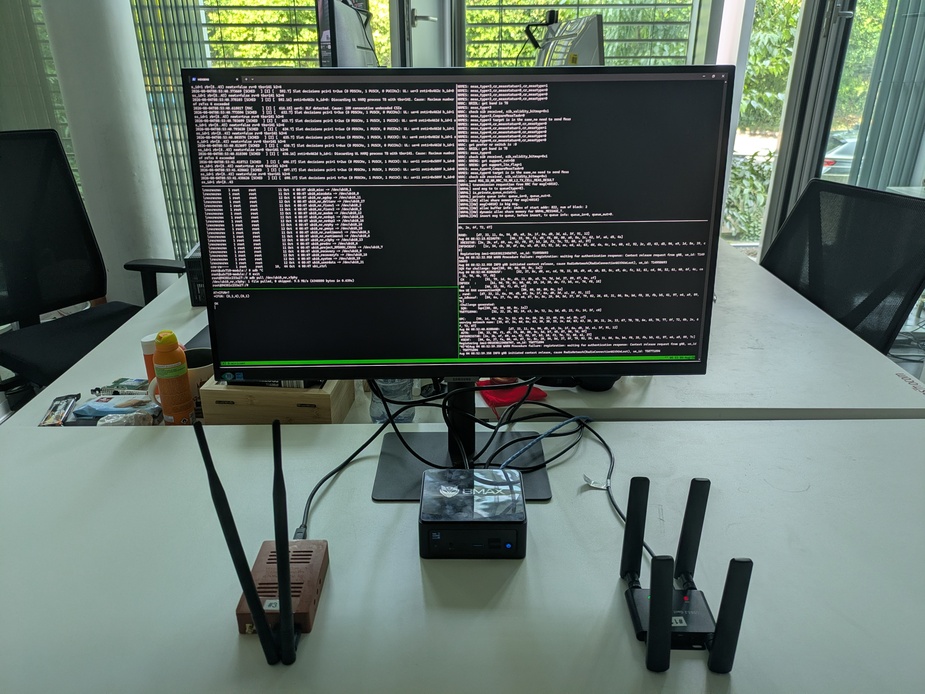}};

\end{tikzpicture}
}
    \caption{
        Experimental testbed for connecting the Quectel RM500U-CNV to a self-provisioned gNB and core network.
        The testbed provides access to gNB, core network and baseband logs; automatically establishes a data session and supports application processor-originated AT commands.
    }
    \label{fig:rm500u-setup}
\end{figure}

\section{Unlocking the Baseband}
\label{sec:foothold}

To analyze the baseband, and ultimately to re-host it, we require access to internal APIs and data structures.
Accessing this information is only possible with code execution inside of the baseband processor. 
To achieve this we unlock the baseband to allow flashing modified baseband firmware. 

\textbf{Application Processor Access.}
The first step is to gain a foothold on the application processor, which on the RM500U-CNV runs a Linux-based firmware.
By default the modem only exposes an AT command interface over USB to the host. We use this interface 
to enable ADB, see \cref{sec:unlock-appendix} for the details.
Afterwards, the modem exposes ADB and provides a root shell on the Linux subsystem.
This gives us our first foothold on the platform and allows us to continue the analysis from within the device.

The root shell obtained through ADB gives us access to the Linux subsystem of the RM500U-CNV.
However, this subsystem is \textit{not} the LTE/5G baseband itself.
Instead, it acts as an orchestration layer for device management, peripheral initialization, firmware loading, 
and coordination of the remaining processors on the platform.

\textbf{Baseband Firmware Loading.}
Having established that Linux orchestrates the remaining co-processors on the platform, we next investigate 
how it loads the LTE/5G baseband firmware.
This orchestration path is necessary because it provides the boundary between the host-facing Linux environment 
and the ThreadX-based baseband subsystem which we intend to analyze.

For the LTE/5G baseband, the \texttt{modem\_control} binary manages the boot process of the modem subsystem.
It loads baseband firmware through a vendor-specific \texttt{procfs} interface exposed by the \texttt{cptl} kernel component.

% writes the firmware image to: \texttt{/proc/cptl/modem} 
% and starts it by writing \texttt{1} to \texttt{/proc/cptl/start}.
% We first analyze the proprietary \texttt{modem\_control} binary, which .
% By reverse engineering this binary, we find that Linux .

Tracing the firmware source further shows that these images live on the Linux side as named UBI volumes, one per component: 
for instance \texttt{ubi0\_nr\_modem} holds the baseband firmware, \texttt{ubi0\_nr\_nrphy} the NRPHY firmware, 
and \texttt{ubi0\_nr\_runtimenv1} the runtime non-volatile data, among more than twenty in total.
% Because they are mounted as UBIFS images on the running device, we dump each one over ADB and analyze it offline, 
% recovering the baseband and co-processor firmware.
% For the baseband, this also gives a concrete loading path for testing whether the platform accepts modified firmware.

% \textbf{Testing Firmware Integrity.}
% To understand whether the platform verifies baseband firmware before execution, we modify one byte in a copy of the baseband image and attempt to boot from the altered firmware.
% The kernel rejects the image and reports the following messages in the \texttt{dmesg} buffer, confirming an implemented integrity check before starting the baseband firmware:

% \begin{center}
%     \begin{minipage}{0.9\linewidth}
%         \begin{lstlisting}
% user.err kernel: [ts] c1 cmp hash_data diffent
% user.err kernel: [ts] c1 verify modem failed
% user.err kernel: [ts] c1 cproc: cptl verify err!
% \end{lstlisting}
%     \end{minipage}
% \end{center}

\textbf{Bypassing Firmware Verification.}
With access to the application processor we can patch \texttt{modem\_control} to load our own modified baseband firmware.
However, the \texttt{ubi0\_nr\_modem} firmware contains a signature which is verified by the \texttt{cptl} kernel component.
The kernel refuses to load incorrectly signed firmware images.
% We reverse engineer the verification path by analyzing the compressed \texttt{vmlinuz} image shipped with the firmware.
We find that the verification logic is \textit{not} backed by a hardware root-of-trust.
The verification therefore depends on software running inside the Linux kernel, where we have \texttt{root} privileges.
% The kernel does not expose convenient debug interfaces such as \texttt{/dev/mem}, which would allow us to patch kernel 
% memory directly at runtime.
% However, the platform does not enforce signature checks for loadable kernel modules.
To bypass the verification, we write a kernel module that patches the verification logic to return success regardless of the 
verification result.

% searches kernel memory for the byte pattern corresponding to 
% the identified verification routine.
% Once located, it resolves the physical page backing the target kernel virtual address, 
% remaps it as writable, and patches the verification logic to return success regardless of the verification result.
% During development, we also observe that the kernel base address is not randomized, and fine-grained KASLR is not available.
% We provide a trace of loading the kernel module below:

% \begin{center}
%     \begin{minipage}{0.98\linewidth}
% \begin{lstlisting}
% root@udx710-module:/ # insmod overwrite-module.ko pattern=fd7bbda90210c0d2

% [ts] c1 hexscan: printk=ffffff80080f29c0
% [ts] c1 hexscan: scanning [ffffff8007e97358 .. ffffff8008297358)
%   < snip >
% [ts] c1 hexscan: writing 30000014 to target address
% [ts] c1 hexscan: MATCH VA=ffffff8008097378 PA=0x7f8088097378
%   < snip >
% [ts] c1 table base: ffffff8008dc4000
% [ts] c1 end of translation reached after 2 steps!
% [ts] c1 *test_pageentry=00e8000084295f13
% [ts] c1 *ver_pageentry=00c0000080200f91
% [ts] c1 flushed.
%   < snip >
% [ts] c1 hexscan: done; matches=1
% \end{lstlisting}
%     \end{minipage}
% \end{center}

% \textbf{Confirming Modified Firmware Execution.}
After loading the kernel module, the platform accepts modified baseband images.
To confirm that the bypass affects the actual firmware executed by the Cortex-R8 subsystem, we modify the baseband version string, reboot the modem, and observe the modified string when querying it via the AT command \texttt{AT+GMR}.
This confirms that we can execute modified LTE/5G baseband firmware and provides the basis for injecting instrumentation.

\section{Introspecting the Baseband}
\label{sec:observing}

To understand what the baseband does, we have to watch it run real procedures.
We build this visibility in layers.
We first drive the modem with a live network so that it performs a real registration, then read what the platform already exposes through its logs, and finally, where the logs fall short, instrument the firmware to inspect and modify execution at runtime.

\textbf{Controlled OTA Baseline.}
Observation is only meaningful once the baseband is doing real work, so we first drive it with a controlled network.
We connect the RM500U-CNV to a local 5G testbed built from an SDR, a software gNB (srsRAN~\cite{srsran_project}), and a core-network backend (open5gs~\cite{open5gs}), and provision a programmable SIM with credentials matching the core so the modem authenticates and registers (cf.~\cref{fig:rm500u-setup}).
This gives us a reproducible, functional registration that we treat as ground truth throughout:
it is the reference against which we validate every firmware modification and, later, the behavior our re-hosted baseband must reproduce.

\textbf{Baseband Logs.}
From the Linux subsystem, the baseband log stream is exposed through \texttt{/dev/slog\_lte}.
Copying from this device file during a normal connection establishment produces tens of megabytes of binary log data.

The stream contains two types of entries:
(i) Plain debug strings or string references that can be resolved directly from the firmware.
These messages are immediately useful for checking whether the modem reaches expected boot and firmware states.
(ii) Tracepoints, where logs contain only a tracepoint identifier and the values of its arguments, while the corresponding format string is stored separately in the vendor's diagnostic tooling.

\textbf{Recovering Tracepoint Strings.}
To decode these tracepoints, we analyze Quectel's diagnostic tooling, Logel, which is distributed through the same forum and support channels used to obtain the firmware.
We find that Quectel Logel retrieves a Windows shared library that contains the mapping between tracepoint identifiers and human-readable format strings.
This library is \textit{publicly accessible} without authentication and contains tracepoint dictionaries for multiple UNISOC-based basebands.
By reverse engineering the library, we recover the tracepoint-to-string mapping for our target and use it to decode the high-level baseband logs from \texttt{/dev/slog\_lte}.

This gives us an extensive view into the running stack and also documents control-plane progress during connection establishment.
We use this mechanism as a convenient way to validate later modifications, including injection of custom debug messages into the baseband firmware.

\textbf{Injecting Probes.}
The logs show which protocol states the baseband reaches, but not how the code produced them, and they give us no way to read arbitrary memory or change state at runtime.
To gain this capability, we add instrumentation to the firmware.
Since we can already run modified images, we insert small shellcode snippets into the Cortex-R8 firmware, compiled with \texttt{arm-none-eabi-gcc} and using \texttt{naked} functions where needed to control the prologue, stack layout, and register preservation.
These probes let us dump memory, print custom values, and modify variables at runtime, but on their own they only expose behavior at manually selected locations and give no general execution feedback.

\textbf{Extending Instrumentation Space.}
Small probes require little space, but injecting extensive tracing instrumentation needs room for trampoline code, metadata, and a reserved trace map.
In addition, we find that appending bytes to the firmware does not make them available at runtime.
By inspecting the firmware header, we find that it encodes the size of the baseband image loaded into memory.
Adjusting this field allows us to expand the loaded firmware image and use the full 32\,MB available in the baseband memory region.

However, not all added memory remains usable after boot, even if successfully loaded.
To identify stable regions, we fill the extended area with a known pattern and dump memory after initialization.
Only about 200 kB remain untouched, as other parts are zeroed out for heap objects, memory pools, and statically allocated data.
We therefore reverse engineer the baseband boot process and identify initialization tables that configure MPU regions and section handlers.
By patching these tables, we reclaim more than 2 MB of memory that remains accessible at runtime and use it for instrumentation code and tracing data.

\textbf{Execution Tracing.}
Hand-placed probes are still too coarse to show how execution flows through the firmware.
As the finest layer of introspection, we add lightweight execution tracing: we record only which call sites execute and record a low-overhead trace of the path the firmware takes that the logs cannot provide.
We implement this tracing by statically rewriting selected call sites in the firmware.

This is challenging since the firmware is proprietary and combines mostly Thumb code with some ARM32 regions. 
Therefore, instruction rewriting first requires reliable mode information.
We use Ghidra scripts to identify candidate call sites together with their execution mode and target addresses in~\cref{tab:coverage-targets}.

\begin{table}[]
  \centering
  \footnotesize
  \begin{tabular}{lr}
    \toprule
    \textbf{Metric} & \textbf{Count} \\
    \midrule
    Basic blocks & 1,031,523 \\
    Branch instructions & 364,527 \\
    Fixed-target call sites & 51,100 \\
    Register-indirect call sites & 12 \\
    \bottomrule
  \end{tabular}
  \caption{Static analysis results used to select execution-tracing targets.}
  \label{tab:coverage-targets}
\end{table}

Full basic-block instrumentation is impractical: instrumenting all 1,031,523 basic blocks would require more than 16 MB of trampoline code, exceeding the 2 MB memory we can safely reclaim.
Instead, we instrument \textit{fixed-target call sites}.
Ghidra identifies 51,100 such sites, which require approximately 817 kB of trampoline code and a 65 kB trace map.
Both fit within the reclaimed memory region.
Each trampoline saves the execution context, calls our instrumentation target, restores the context, and then branches to the original call target (\cref{fig:trampoline}).
The instrumentation target itself checks a runtime flag before writing trace entries, which lets us enable collection only after the modem has completed initialization (\cref{fig:cov-hit}); we detail both in \cref{sec:tracing-appendix}.

At runtime, the execution-trace map is stored in the reclaimed baseband memory region.
From the Linux subsystem, we dump this region through the existing \texttt{/proc/cptl/modem} interface by seeking to the trace-coverage-map address and reading the corresponding memory range.
This allows us to collect execution traces between runs without rebooting the full device.

\textbf{Protocol-Level Instrumentation.}
Execution traces tell us which code is executed, but not which protocol messages caused it.
We therefore add probes around RRC transmission and reception paths.
These probes dump incoming and outgoing payloads, queue items exchanged between tasks, and messages passed between RRC and lower layers.
By correlating these dumps with execution-trace feedback, we connect protocol-level behavior to internal execution paths.

We also find the functions that update protocol state; by hooking them we can watch how the state changes, a capability that is useful for testing.

\textbf{Interactive Debug Task.}
Beyond static probes and tracing, we also add a small interactive debug task to the ThreadX firmware. By reverse engineering the RTOS task-creation routines, we find that the firmware allows new tasks to be spawned at runtime. We use this mechanism to create a custom debug task during boot and provide it with its own stack and command handler.

The task exposes a simple switch-based interface for runtime experiments. It supports commands such as memory peek and poke, controlled assertions, packet injection, tracing control, and calls to selected firmware functions with custom arguments. This gives us a practical way to explore firmware behavior interactively without rebuilding the firmware image for every small experiment.

This task complements the static instrumentation pipeline: the trampolines provide systematic execution tracing, while the debug task provides an interactive control channel for experiments.

\section{Reconstructing the Peripheral Environment}
\label{sec:periph}
The instrumentation introduced above lets us place a probe at any point in the firmware, in particular at the boundaries between SoC processors.
Combined with the testbed, which drives real traffic across the $\{$AP, NRPHY, SIM, SEC Engine$\}$ $\leftrightarrow$ baseband boundaries, this exposes actual peer behavior.

Applying static analysis to co-processor firmware images alone to recover peer behavior is generally infeasible: corresponding logic is distributed across images or implemented partly in hardware.
It can only be recovered by observing execution during runtime.
We use this runtime behavior to characterize each peer the baseband depends on, which we reconstruct as a model in the emulator.
The resulting architecture, shown in~\cref{fig:platform-arch}, focuses on the components relevant to 5G operation and omits legacy subsystems, such as GSM and UMTS processing, which are outside the scope of this analysis.

\textbf{Overview.} The Linux subsystem runs on two ARM Cortex-A55 cores and provides the host-facing management environment.
The LTE/5G protocol stack executes separately on two ARM Cortex-R8 cores running Microsoft ThreadX.
Additional co-processors implement lower-layer functionality, including PHY and MAC processing.
Our actual analysis target is the ThreadX-based baseband firmware executing on the Cortex-R8 subsystem, while the Linux subsystem serves as the entry point for controlling and observing it.

\begin{figure}
\centering
\resizebox{\linewidth}{!}{
\begin{tikzpicture}[
  font=\footnotesize,
  ap/.style={draw=black!55, fill=black!8, drop shadow, align=center,
             minimum width=2.8cm, minimum height=0.85cm},
  bb/.style={draw=black!70, thick, drop shadow, align=center,
             minimum width=4.4cm, minimum height=0.9cm, fill=cFW!22},
  fw/.style={draw=black!55, fill=cFW!22, drop shadow, align=center,
             minimum width=1.7cm, minimum height=0.82cm, font=\scriptsize},
  sim/.style={fw, fill=cSIM!24},
  hw/.style={fw, fill=cHW!20},
  clus/.style={draw=black!35, dash pattern=on 2pt off 1.5pt, rounded corners=2pt, inner sep=3pt},
  load/.style={-{Stealth[length=2.4mm]}, line width=1.1pt, black!80},
  wire/.style={black!45},
]
\definecolor{cFW}{HTML}{F2C230}
\definecolor{cSIM}{HTML}{E38B2A}
\definecolor{cHW}{HTML}{3F7FD6}
\newcommand{\archstar}[3]{\node[star, star points=5, star point ratio=2.3, fill=#2,
    draw=black!55, line width=0.2pt, minimum size=3mm, inner sep=0] at (#1.north east) (#3) {};}
\newcommand{\archleg}[1]{\tikz\node[star,star points=5,star point ratio=2.3,fill=#1,draw=black!55,minimum size=2.3mm,inner sep=0]{};}

\node[ap] (ap) at (-0.6,3.0) {\textbf{Application Processor}\\[-1pt]{\scriptsize 2$\times$ Cortex-A55 $\cdot$ Linux}};
\node[bb] (bb) at (-0.6,1.5) {\textbf{Baseband Processor}\\[-1pt]{\scriptsize 2$\times$ Cortex-R8 $\cdot$ ThreadX}};
\draw[load] (ap.south) -- node[right, font=\scriptsize]{loads firmware, drives} (bb.north);
\archstar{bb}{cFW}{bbs}

\node[fw] (nrphy) at (-3.7,0.15) {NRPHY\\{\scriptsize 2$\times$ Cortex-R5}};
\node[fw] (nrdsp) at (-3.7,-0.85){NR DSP\\{\scriptsize CEVA-XC4500}};
\archstar{nrphy}{cFW}{nrphys}\archstar{nrdsp}{cFW}{nrdsps}
\node[clus, fit=(nrphy)(nrdsp), label={[font=\scriptsize\itshape,black!55]below:NR L1/L2}] (nrstuff) {};

\node[fw]  (nv)  at (2.55,0.15) {NVram};                       \archstar{nv}{cFW}{nvs}
\node[sim] (sim) at (2.55,-0.85) {SIM / UICC};                 \archstar{sim}{cSIM}{sims}

\node[hw] (sec)  at (-0.6,0.15) {Crypto\\{\scriptsize SEC Engine}};
\node[hw] (sipa) at (-1.6,-0.85){IP accelerator \\{\scriptsize (SIPA)}};
\node[hw] (clk)  at (0.4,-0.85){Clock};
\archstar{sec}{cHW}{secs}\archstar{sipa}{cHW}{sipas}\archstar{clk}{cHW}{clks}
\node[clus, fit=(sec)(sipa)(clk), label={[font=\scriptsize\itshape,black!55]below:Hardware}] (hwstuff) {};
\node[clus, fit=(nv)(sim), label={[font=\scriptsize\itshape,black!55]below:Configuration}] (configstuff) {};

\foreach \n in {configstuff,hwstuff,nrstuff}{ \draw[wire] (bb.south) -- (\n.north); }

\node[draw=black!45, rounded corners=2pt, fill=white, drop shadow, align=left,
      font=\scriptsize, anchor=north west, inner sep=4pt] at (-5.00,3.45) {%
  \archleg{cFW}~ Firmware (loaded by AP)\\[2pt]
  \archleg{cSIM}~ SIM (pre-programmed)\\[2pt]
  \archleg{cHW}~ Silicon-only Hardware};
\end{tikzpicture}
}
    \caption{Reverse-engineered architecture of the RM500U-CNV. The application processor loads and drives the baseband, which coordinates the co-processors and peripherals. Stars mark what each component is: firmware loaded by the AP, the pre-programmed SIM, or fixed hardware.}
\label{fig:platform-arch}
\end{figure}
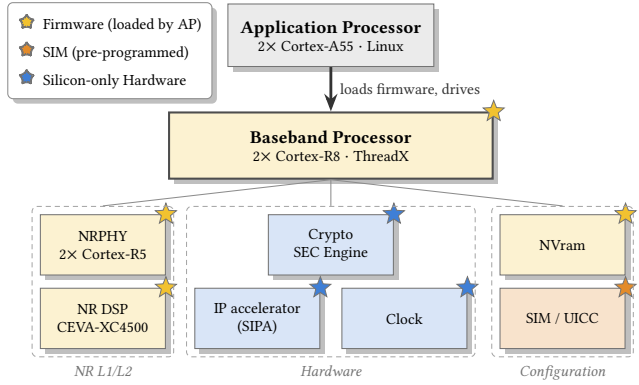

\begin{figure}
\centering
\begin{tikzpicture}[
    >=Stealth,
    node distance=2.2cm,
    comp/.style={
        draw,
        rounded corners,
        minimum width=2.8cm,
        minimum height=1.3cm,
        align=center,
        fill=white,
        drop shadow
    },
    msg/.style={
        font=\small,
        midway,
        fill=white,
        inner sep=1pt,
    }
]

\node[comp] (bb) {Baseband\\(Peer HAL Task)};
\node[comp, right=of bb] (nrphy) {SoC Peer};

\draw[->, thick]
    ($(bb.east) + (0,0.2)$) -- ($(nrphy.west) + (0,0.2)$)
    node[msg, above] {(1) \texttt{REQ} / (2) \texttt{IND}};

\draw[<-, thick]
    ($(bb.east) + (0,-0.2)$) -- ($(nrphy.west) + (0,-0.2)$)
    node[msg, below] {(4) \texttt{CNF} / (3) \texttt{RSP}};

\node[comp, below= 1.0cm of bb] (bb2) {Baseband\\(Peer HAL Task)};
\node[comp, right=of bb2] (nrphy2) {SoC Peer};

\node[below=0.2cm] (sol) at ($(bb.south)!0.5!(nrphy.south)$) {Solicited Communication};
\node[below=0.2cm] (unsol) at ($(bb2.south)!0.5!(nrphy2.south)$) {Unsolicited Communication};

\draw[<-, thick]
    ($(bb2.east) + (0,0.2)$) -- ($(nrphy2.west) + (0,0.2)$)
    node[msg, above] {(2) \texttt{IND} / (1) \texttt{REQ}};

\draw[->, thick]
    ($(bb2.east) + (0,-0.2)$) -- ($(nrphy2.west) + (0,-0.2)$)
    node[msg, below] {(3) \texttt{RSP} / (4) \texttt{CNF}};

\end{tikzpicture}

    \caption{Communication modes between the baseband and the SoC peers. Solicited communication is baseband-originated, e.g., SIM requests, while unsolicited communication is originated by an SoC component, e.g., the AP.}
    \label{fig:comm-model}
\end{figure}
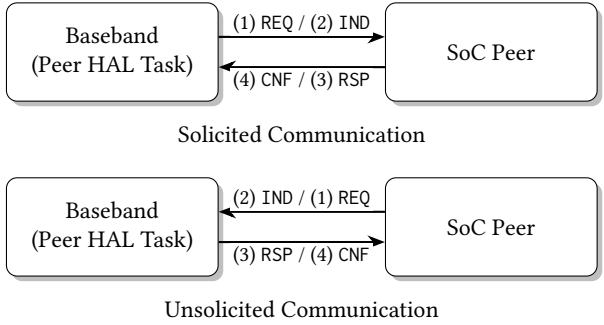

\textbf{Communication Model.}
Most peers exchange messages with the baseband asynchronously, over shared-memory ring buffers and inter-processor interrupts, in one of two patterns (cf.~\cref{fig:comm-model}).
In a \emph{solicited} exchange, the baseband issues a request and the peer returns the matching response.
In an \emph{unsolicited} exchange, the peer raises an indication on its own, and the baseband replies.
Most peers use both directions, so we describe each one below in terms of the messages it exchanges under both patterns.

\textbf{Application Processor.}
The application processor drives the baseband over two channels.
The first carries high-level AT commands, which the baseband parses and answers.
The second operates at a lower level and is more difficult to reconstruct: over the \texttt{smsg}/\texttt{sblock} control channel, the AP configures a data-plane Ethernet interface between the AP kernel and the baseband, set up together with the SIPA hardware module that accelerates the IP data path.
Both channels mix solicited and unsolicited messages: the AP issues unsolicited commands, for example to force registration or attach, and in turn receives unsolicited messages from the baseband, such as notifications that the SIM is ready or that the baseband has come up.

\textbf{NRPHY Co-Processor.}
The NRPHY co-processor runs the lower layers of the 5G stack (PHY and MAC).
The baseband communicates with it through the NRPAL task over shared-memory ring buffers and inter-processor interrupts, using both patterns: a solicited request is delivered to the NRPHY as an indication and answered with a response the baseband receives as a confirmation, while an unsolicited indication from the NRPHY, for example a synchronization event or a received transport block, is answered by the baseband.

\textbf{SIM Card.}
The SIM differs qualitatively from the other peers.
It is not only storage for secrets: it is a separate secure processor running Java applets~\cite{lisowski2024simurai}, performing key derivation and holding non-volatile state, reached by the baseband over the standardized APDU protocol.
Unlike the other peers, which are asynchronous and signal completion by interrupt, the SIM is driven synchronously: the baseband issues an APDU and blocks on the reply; if the card is unreachable or does not answer in time, the baseband gives up and assumes no SIM is present.
A working SIM is therefore a precondition for cell attach and authentication, the two steps that gate all further baseband operation.
Reconstructing this peer means standing in for the applet and its APDU exchange.

\textbf{Crypto Co-Processor.}
The crypto co-processor is implemented entirely in hardware, so unlike the other peers there is no firmware to reverse engineer.
The algorithms it runs are standard and well documented, but its interface to the baseband, the registers and DMA channels it exposes, is undocumented and can only be recovered by observation on real silicon.
It is nonetheless essential: it verifies the authenticity and integrity of protected messages, and the baseband cannot proceed past security-mode setup without it.

\textbf{NVRAM.}
Not every environmental dependency communicates with the baseband as a peer.
The non-volatile memory (NVRAM), shipped as multiple baseband memory-mapped UBIFS volumes, holds RF calibration data and configuration parameters and is mapped directly into the baseband's address space rather than accessed through request/response messages.
Yet it is critical: if it is absent, or mapped at the wrong location, the baseband fails to boot.
Dumping the image itself is trivial, as it can be extracted like any other UBI volume; what matters for reconstruction is placing it at the address and offset the firmware expects.

Together with the clock, these peers and the NVRAM configuration constitute the environment the baseband depends on to execute.
Having characterized what each one does and when, we can now reconstruct each peer as a model, map the NVRAM image at the address the firmware expects, and drive the firmware against these models deterministically, in an emulator (\cref{sec:rehost}).

\section{Deterministic Baseband Re-Hosting}
\label{sec:rehost}

On-device instrumentation makes the baseband observable, but it is not practical to drive at scale:
(i) every experiment requires a full SoC reboot with at least 60 seconds of execution time.
(ii) Additionally, driving even a live procedure needs a provisioned SIM and a full radio testbed rather than a single command; and the probes recovered in \cref{sec:observing} were themselves expensive to build, so adding a new one for every question does not scale.
(iii) Finally, tests can only be carried out on-device, impacting horizontal scaling of experiments.

We therefore postulate the same access without the on-device cost: we run the same firmware off the device, under a virtual clock we control, so that resets are cheap, instrumentation is trivial to add, and every run reproduces.
Re-hosting is the final step of the chain this case study has followed:
gaining code execution, running modified firmware, adding instrumentation, and recovering the peer models of \cref{sec:periph} now let us run the firmware against those models deterministically.

\subsection{Peer- and Interconnect-Driven Re-Hosting}

We define a QEMU machine for the baseband's Cortex-R8 subsystem, constrained to the same rules as real silicon: the firmware may only be driven through the SoC interconnect and peripherals, i.e., DMA, message rings, mailboxes, interrupts, MMIO, and APDU (cf.~\cref{sec:background}).
We never hook arbitrary functions or write to memory the firmware does not already share with a peer over that interconnect.
Every component so introduced is characterized peer by peer in \cref{sec:periph}.

Consequently, emulation fidelity rests on \textit{two properties} that do not always coincide:
(i) fidelity to a \textit{peer's behavior}, and (ii) fidelity to the \textit{interconnect} it uses to express that behavior.
We illustrate both dimensions with an example:
the NRPHY co-processor does \textit{not} signal a pending downlink message over a GIC interrupt line;
it rings a fixed inter-core mailbox doorbell, which only the baseband firmware's mailbox HAL observes and never re-raises as a GIC interrupt.
A model that injected a GIC IRQ instead, assuming any interrupt mechanism would do, would capture the peer's intent but miss its interconnect, and the firmware would never reach the intended downlink handler.

We refer to this abstraction as \emph{peer- and interconnect-driven} re-hosting:
we reconstruct every peer the firmware depends on from every introspection point available, static and dynamic, and drive the firmware only through the exact channels a real peer would use, rather than re-hosting the firmware in isolation and patching around whatever is missing.

\subsection{Faithful Re-Hosting Requirements}
Building a faithful re-host this way rests on five requirements:
  (i) a real device available for instrumentation and live interconnect probing;
  (ii) a controlled radio testbed, to drive the baseband through real network procedures;
  (iii) the firmware images of the baseband and every co-processor it depends on, for static analysis;
  (iv) visibility into the Linux kernel's radio-interface-layer driver, the only source of AP-originated commands; and
  (v) the emulator itself: a QEMU machine for the baseband core, together with a model for every peer it depends on.
Together, these give us every introspection point needed to reconstruct each peer as a model.

Existing baseband re-hosts rely mainly on (iii) and (v), a firmware image and an emulator core.
Treating (i) as live probing, as opposed to memory-snapshotting~\cite{klischies2025basebridge}, and adding (ii) and (iv), is what faithful, peer- and interconnect-driven re-hosting additionally requires.

\subsection{Re-Hosting Methodology}
The interconnect fixes a clear abstraction:
a peer is fully described by the MMIO regions, DMA ranges, doorbells, and interrupts it exposes, and the firmware's behavior is entirely determined by what it reads and writes on the interconnect.
Therefore, a faithful model has to respect this boundary rather than the firmware's internal state.

We first identify addresses, IRQs, and doorbells, (i) statically from a firmware image or the kernel driver where one is possible, and (ii) dynamically by probing the real device where runtime information is required.
Once located, we inspect the values egressed and ingressed at these call and read/write sites.
An execution trace allows inferring which code region to instrument:
an interrupt service routine, or the code that reads a specific memory address.
Probing exactly those points lets us formalize a peer's message formats and firing conditions from introspection rather than static assumption.
This order, (i) locate the addresses, (ii) trace the execution, (iii) then probe the sites it points to, is the same for every peer.

In practice, fidelity follows directly from what the testbed lets us drive and inspect, not from an a priori scope decision: we model only what we can exercise.
Modeling another co-processor is therefore not free; it requires extending the testbed to reach it.
LTE PHY, for instance, is driven over a different channel than NRPHY:
modeling it first requires driving the baseband down that channel to have anything to inspect~\cite{klischies2025basebridge}.

\subsection{Determinism}

Execution determinism is a key requirement for re-hosting: peer fidelity, interconnect fidelity, and reproducibility ultimately rely on whether execution follows the same virtual timeline real silicon would, not only whether the same events eventually occur.
Protocol timers show this distinction: the cellular control plane is timer-driven throughout.
Guard timers, supervision timers, and retransmission timers release a connection or trigger a state transition once no response arrives in time, and each one only fires, or stays silent, if the firmware's notion of elapsed time tracks real hardware.
If timers are modeled too loosely, events either fire early and abort a procedure that would have completed on real silicon, or never fire at all, silently dropping behavior the evaluation depends on.

We therefore derive the firmware's notion of time from its own execution, not the host clock.
We run the guest under QEMU's \texttt{icount} mode at \texttt{shift=3}, which ties virtual time to the guest's retired instruction count:
one ThreadX RTOS timer tick, the smallest interval the firmware's scheduler observes, corresponds to 125{,}000 instructions, i.e., 1\,ms of virtual time, regardless of how long that takes to compute on the host.
The difference is measurable: under plain QEMU TCG, the periodic tick preempts a real-time-sensitive routine, for instance the SIM bring-up walk, at a host-jitter-dependent point on every run.
Its depth and reset count subsequently vary run to run and introduce non-determinism.
Under \texttt{icount}, two independent runs of the same routine produce a byte-for-byte identical trace.

Instruction-counted time makes the firmware itself deterministic; the only source of non-determinism left is when, and what, a peer injects.
We hold peers to the same requirements:
every peer input is delivered only at a tick boundary the firmware's own clock has reached, never on a wall-clock schedule, a retry, or decoded log output, so that timer-dependent behavior on both sides of the interconnect tracks one virtual timeline.

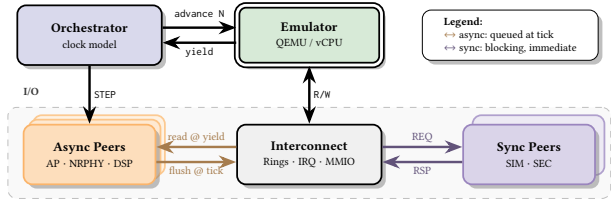
\begin{figure}[t]
  \centering
  \resizebox{0.95\linewidth}{!}{% Co-simulation architecture, interaction view (from whiteboard architecture_cosim.jpg).
% Same color coding as the steady-state sequence diagram (fig:determinism):
% Orchestrator (navy) / Async Peers (orange) / Emulator (green) / Sync Peers (purple).
% Transposed to a 2-row x 3-column grid so the figure is wide and short rather
% than narrow and tall: top row is the control pair (Orchestrator, Emulator,
% legend); bottom row is the peer/interconnect plane (Async Peers,
% Interconnect, Sync Peers), tied to the row above by short vertical edges.
\begin{tikzpicture}[
    >=Stealth,
    orch/.style={draw, rounded corners, thick, drop shadow, fill=NavyBlue!15,
                 minimum width=2.3cm, minimum height=0.95cm, font=\scriptsize\bfseries, align=center},
    emu/.style={draw, double, double distance=1.3pt, rounded corners, thick, drop shadow, fill=ForestGreen!18,
                minimum width=2.3cm, minimum height=0.95cm, font=\scriptsize\bfseries, align=center},
    ic/.style={draw, rounded corners, thick, drop shadow, fill=gray!12,
               minimum width=2.3cm, minimum height=0.95cm, font=\scriptsize\bfseries, align=center},
    edge/.style={<->, line width=1.1pt},
    lbl/.style={font=\tiny, align=center, inner sep=1.5pt},
    grouplbl/.style={font=\tiny\bfseries},
]

\def\xL{0}    % left column: Orchestrator / Async Peers
\def\xM{3.5}  % middle column: Emulator / Interconnect
\def\xR{7.0}  % right column: Legend / Sync Peers
\def\yTop{0.95}
\def\yBot{-0.95}

% ---- top row: control pair + legend ----
\node[orch] (orch) at (\xL,\yTop) {Orchestrator\\{\tiny\normalfont clock model}};
\node[emu]  (emu)  at (\xM,\yTop) {Emulator\\{\tiny\normalfont QEMU / vCPU}};
\node[draw, rounded corners, fill=white, drop shadow, align=left, font=\tiny,
      minimum width=2.8cm, minimum height=0.8cm, anchor=center] (legend) at (\xR-0.3,\yTop) {
  \textbf{Legend:}\\
  \textcolor{BurntOrange!60!black}{$\leftrightarrow$} async: queued at tick\\
  \textcolor{RoyalPurple!55!black}{$\leftrightarrow$} sync: blocking, immediate
};

% ---- bottom row: peer / interconnect plane, boxed as the I/O boundary ----
\begin{pgfonlayer}{background}
  \node[draw=gray!50, dashed, rounded corners, fill=gray!5,
        fit={(\xL-1.3,\yBot-0.7) (\xR+1.3,\yBot+0.75)}, inner sep=0] (iobox) {};
\end{pgfonlayer}
\node[font=\tiny\bfseries, gray!45!black, anchor=south west] at ([xshift=4pt,yshift=2pt]iobox.north west) {I/O};

% Async Peers: compact overlapping stack, label + subtitle INSIDE the front card
\node[draw=BurntOrange!45, rounded corners, thick, fill=BurntOrange!10,
      minimum width=2.1cm, minimum height=0.95cm] at (\xL+0.15,\yBot+0.15) {};
\node[draw=BurntOrange!50, rounded corners, thick, fill=BurntOrange!17,
      minimum width=2.1cm, minimum height=0.95cm] at (\xL+0.08,\yBot+0.08) {};
\node[draw=BurntOrange!65, rounded corners, thick, drop shadow, fill=BurntOrange!26,
      minimum width=2.1cm, minimum height=0.95cm, font=\scriptsize\bfseries, align=center]
      (async) at (\xL,\yBot) {Async Peers\\{\tiny\normalfont AP $\cdot$ NRPHY $\cdot$ DSP}};

\node[ic] (ic) at (\xM,\yBot) {Interconnect\\{\tiny\normalfont Rings $\cdot$ IRQ $\cdot$ MMIO}};

% Sync Peers: compact overlapping stack, label + subtitle INSIDE the front card
\node[draw=RoyalPurple!40, rounded corners, thick, fill=RoyalPurple!8,
      minimum width=2.1cm, minimum height=0.95cm] at (\xR+0.15,\yBot+0.15) {};
\node[draw=RoyalPurple!55, rounded corners, thick, drop shadow, fill=RoyalPurple!16,
      minimum width=2.1cm, minimum height=0.95cm, font=\scriptsize\bfseries, align=center]
      (sync) at (\xR,\yBot) {Sync Peers\\{\tiny\normalfont SIM $\cdot$ SEC}};

% ---- edges: the credit protocol, top row (Orchestrator <-> Emulator) ----
\draw[->, line width=1.1pt] ([yshift=0.12cm]orch.east) -- node[lbl,above,font=\tiny\bfseries\tt,yshift=0.1cm]{advance N} ([yshift=0.12cm]emu.west);
\draw[->, line width=1.1pt] ([yshift=-0.12cm]emu.west) -- node[lbl,below,font=\tiny\bfseries\tt]{yield} ([yshift=-0.12cm]orch.east);

% ---- edge: left column, Orchestrator steps every async peer directly ----
\draw[->, line width=1.1pt] (orch.south) -- node[lbl,right,font=\tiny\bfseries\tt]{STEP} (async.north);

% ---- edge: middle column, Emulator only ever touches the Interconnect ----
\draw[edge] (emu.south) -- node[lbl,right,font=\tiny\bfseries\tt]{R/W} (ic.north);

% ---- edge: bottom row left, Interconnect <-> Async Peers, flushed at tick boundary ----
\draw[edge, BurntOrange!60!black, <-] ([yshift=0.12cm]async.east) -- node[lbl,above]{read @ yield} ([yshift=0.12cm]ic.west);
\draw[edge, BurntOrange!60!black, ->] ([yshift=-0.12cm]async.east) -- node[lbl,below]{flush @ tick} ([yshift=-0.12cm]ic.west);

% ---- edge: bottom row right, Interconnect <-> Sync Peers ----
\draw[->, line width=1.1pt, RoyalPurple!55!black] ([yshift=0.12cm]ic.east) -- node[lbl,above]{REQ} ([yshift=0.12cm]sync.west);
\draw[<-, line width=1.1pt, RoyalPurple!55!black] ([yshift=-0.12cm]ic.east) -- node[lbl,below]{RSP} ([yshift=-0.12cm]sync.west);

\end{tikzpicture}}
  \caption{The co-emulation architecture. The orchestrator drives the emulator on a strict \texttt{advance}/yield credit protocol and steps every async peer directly; the emulator itself never addresses a peer, it only reads and writes the interconnect, the modeled hardware mechanism. The interconnect queues async-peer I/O and applies it only at the next tick boundary; sync peers are instead reached through a direct blocking callback that returns before the tick ends.}
  \label{fig:architecture}
\end{figure}

\textbf{Execution Model.}
\cref{fig:architecture} lays out the components this model coordinates.
An \textit{orchestrator} owns virtual time; the \textit{emulator} runs the guest and never addresses a peer directly, only the modeled \textit{interconnect}; and the peers on the other side of that interconnect split into the two channel kinds already mapped out there: \textit{asynchronous peers} (AP, NRPHY, DSP), whose input the interconnect only applies at a tick boundary, and \textit{synchronous peers} (SIM, SEC), reached through a direct blocking call that returns before the tick ends.

The orchestrator drives the emulator in bounded units of virtual time called \textit{quantums}.
Granting one is a single command, \texttt{advance N}: it hands the emulator $N$ tick credits, one credit per periodic timer interrupt, so the firmware is authorized to advance exactly $N$ ThreadX scheduler iterations, before the orchestrator regains control.

A quantum ends, and control returns to the orchestrator, at exactly one of two moments: either the last of the $N$ credits drains, meaning the firmware ran the full quantum, or the firmware goes idle (via \texttt{WFI}) with credits still unused.
Either way the emulator reports this boundary back; the orchestrator injects asynchronous peer events, waits for the firmware to acknowledge them, and only then grants the next quantum before repeating.
A graphical representation of the initialization and steady-state co-emulation of peers and the emulator is given in \cref{fig:determinism}.

The orchestrator is not only waiting for control to return to it; it chooses the size of every quantum to be the \textit{exact distance} to an unsolicited peer event.
Therefore, the boundary at which it regains control is always self-chosen.
Every scheduled action is therefore a function of virtual time or of a modeled hardware event the firmware itself produced.
Importantly, we forbid pacing by wall-clock sleeps, retransmission on timeout, and synchronizing on a decoded log line, since all three make the outcome depend on host scheduling rather than on the guest's own state.

\textbf{Deterministic Snapshots.}
Determinism yields a second benefit beyond reproducibility: a copy-on-write \texttt{fork()} is valid at any point in the run, not only at boot.
This holds because every peer, not only the firmware, advances exclusively on the same instruction-counted grid and never reads wall-clock state.
We exploit this property directly in the implementation: each peer is a non-blocking step function, one that drains whatever input is ready, reacts, and returns, rather than a loop that owns its own thread, and the same per-tick callback that advances the virtual clock drives every peer's step in turn.
At any instant, therefore, exactly one thread of control exists in the whole process, guest and peers alike, which is precisely what a safe fork requires: taken at a quantum boundary on that thread, it captures the guest CPU, guest RAM, and every peer's state together, as a single image.

\begin{figure}[t]
  \centering
  \resizebox{0.95\linewidth}{!}{% Steady-state co-simulation loop (from whiteboard seq_diag.jpg). Lifelines:
% Orchestrator (barrier / time-keeper) sits flush on the left edge of the
% drawing; Sync Peers (a compact stack: SIM, SEC) sits flush on the right
% edge, reached by a direct blocking call that answers immediately, entirely
% outside the credit/tick loop. Interpreter (a compact stack: AP, NRPHY, DSP,
% each a non-blocking step driven by the same per-tick pump) and Machine
% (QEMU / vCPU) sit in the two interior columns.
\begin{tikzpicture}[
    >=Stealth,
    actor/.style={draw, rounded corners, thick, drop shadow, fill=NavyBlue!15,
                  minimum width=2.6cm, minimum height=1.1cm, font=\small\bfseries,
                  align=center},
    life/.style={draw, gray!55, thick},
    msg/.style={->, line width=1.15pt},
    lbl/.style={font=\small, align=center, fill=NavyBlue!6, inner sep=1.5pt},
    lblbox/.style={font=\small, align=center, fill=gray!11, inner sep=1.5pt},
    note/.style={font=\scriptsize\itshape, gray!35!black, align=left, inner sep=1pt},
    phaselbl/.style={font=\scriptsize\bfseries, gray!25!black},
    annobox/.style={fill=gray!11, draw=gray!45, rounded corners, dashed},
    annolbl/.style={font=\small\bfseries, gray!45!black},
]

\def\xA{0}     % Orchestrator -- flush left edge
\def\xI{3.9}   % Interpreter (stack: AP, NRPHY, DSP)
\def\xM{7.6}   % Machine
\def\xS{11.4}  % Sync Peers (stack: SIM, SEC) -- flush right edge
\def\ytop{0}
\def\ybot{-15.5}

% ---- Orchestrator (anchored so its lifeline sits on the box's left edge) ----
\node[actor, anchor=west] (A0) at (\xA,\ytop) {Orchestrator\\{\scriptsize\normalfont barrier\,/\,time-keeper}};
% ---- Machine, centered in its interior column ----
\node[actor, fill=ForestGreen!18] (M0) at (\xM,\ytop) {Machine\\{\scriptsize\normalfont QEMU / vCPU}};

% ---- Interpreter: a compact overlapping stack, label INSIDE the front card ----
\node[draw=BurntOrange!45, rounded corners, thick, fill=BurntOrange!10,
      minimum width=2.0cm, minimum height=1.1cm] at (\xI+0.15,\ytop+0.15) {};
\node[draw=BurntOrange!50, rounded corners, thick, fill=BurntOrange!17,
      minimum width=2.0cm, minimum height=1.1cm] at (\xI+0.08,\ytop+0.08) {};
\node[draw=BurntOrange!65, rounded corners, thick, drop shadow, fill=BurntOrange!26,
      minimum width=2.0cm, minimum height=1.1cm, font=\small\bfseries, align=center]
      (IStack) at (\xI,\ytop) {Interpreter\\{\scriptsize\normalfont AP $\cdot$ NRPHY $\cdot$ DSP}};

% ---- Sync Peers: a compact overlapping stack, flush on the right edge ----
\node[draw=RoyalPurple!40, rounded corners, thick, fill=RoyalPurple!8,
      minimum width=1.9cm, minimum height=1.1cm, anchor=east] at (\xS-0.15,\ytop+0.15) {};
\node[draw=RoyalPurple!55, rounded corners, thick, drop shadow, fill=RoyalPurple!16,
      minimum width=1.9cm, minimum height=1.1cm, anchor=east, font=\small\bfseries, align=center]
      (SStack) at (\xS,\ytop) {Sync Peers\\{\scriptsize\normalfont SIM $\cdot$ SEC}};

% ---- lifelines (all start right below the boxes; no external captions to clear) ----
\draw[life, dashed] (\xA,\ytop-0.65) -- (\xA,\ybot);
\draw[life, dashed] (\xI,\ytop-0.65) -- (\xI,\ybot);
\draw[life, dashed] (\xM,\ytop-0.65) -- (\xM,\ybot);
\draw[life, dashed] (\xS,\ytop-0.65) -- (\xS,\ybot);

% ---- phase background bands (flush to the outer lifelines, minimal padding) ----
\begin{pgfonlayer}{background}
  \node[draw=ForestGreen!45, rounded corners, fill=ForestGreen!5,
        fit={(\xA-0.25,-1.0) (\xS+0.25,-5.2)}, inner sep=0] (initband) {};
  \node[draw=NavyBlue!45, rounded corners, fill=NavyBlue!5,
        fit={(\xA-0.25,-5.55) (\xS+0.25,-13.1)}, inner sep=0] (steadyband) {};
  \node[fit={(\xA-0.25,-13.45) (\xS+0.25,\ybot)}, inner sep=0] (steadyband2fit) {};
  \shade[top color=NavyBlue!12, bottom color=white]
        (steadyband2fit.north west) rectangle (steadyband2fit.south east);
  \draw[NavyBlue!45, rounded corners]
        (steadyband2fit.north west) rectangle (steadyband2fit.south east);
\end{pgfonlayer}
\node[phaselbl, ForestGreen!40!black, anchor=north east, yshift=-2mm, xshift=-1mm] at (initband.north east) {\large \;INITIALIZATION~~~};
\node[phaselbl, NavyBlue, anchor=north east, yshift=-2mm, xshift=-1mm] at (steadyband.north east) {\large\;STEADY STATE\;~~~};
\node[phaselbl, NavyBlue!45!black, anchor=north east, yshift=-2mm, xshift=-1mm] at (steadyband2fit.north east) {\large\;NEXT QUANTUM~~~};

% ==================== INITIALISATION ====================
\begin{scope}[lbl/.append style={fill=ForestGreen!6}]
\draw[msg] (\xA,-1.7) -- node[lbl,above]{Spawn Peers} (\xI,-1.7);
\draw[msg] (\xA,-2.4) -- node[lbl,above]{Create Machine Object} (\xM,-2.4);
\draw[msg] (\xM,-3.1) -- node[lbl,above]{Attach to Peers} (\xI,-3.1);
\draw[msg] (\xI,-3.8) -- node[lbl,above]{Attach OK} (\xM,-3.8);
\draw[msg] (\xM,-4.5) -- node[lbl,above]{Machine Ready} (\xA,-4.5);
\end{scope}

% ==================== STEADY (loop) ====================
\draw[msg] (\xA,-6.25) -- node[lbl,above,font=\tt\small\bfseries]{advance N} node[lbl,below]{grant ticks (credit)} (\xM,-6.25);

% pseudocode: what the machine does with the granted credit, top right of the gap
\node[note, anchor=north east, font=\ttfamily\scriptsize] at (\xS-0.25,-6.45)
     {WHILE credits > 0:\\\quad credits -= 1;\\\quad EXEC TICK;};

% ---- meanwhile: sync peers answer directly, boxed off from the tick loop ----
\draw[msg] (\xM,-8.55) -- node[lblbox,above]{MMIO / APDU write $\cdot$ read} (\xS,-8.55);
\draw[msg] (\xS,-8.85) -- node[lblbox,below]{IRQ} (\xM,-8.85);
\begin{pgfonlayer}{background}
      \node[annobox, fit={(\xM-0.0,-7.7) (\xS+0.1,-9.2)}, inner sep=3pt] (syncbox) {};
\end{pgfonlayer}
\node[annolbl, anchor=north west] at ([xshift=3pt,yshift=0pt]syncbox.north west) {Synchronous Communication};

\draw[msg] (\xM,-9.75) -- node[lbl,above,font=\tt\bfseries]{ADVERTISE} node[lbl,below,font=\tt]{credit == 0} (\xA,-9.75);

\draw[msg] (\xA,-10.55) -- node[lbl,above,font=\tt\bfseries]{STEP} node[lbl,below]{drive every peer's non-blocking step} (\xI,-10.55);

\draw[msg] (\xI,-11.25) -- node[lbl,above]{write to interconnect} (\xM,-11.25);

\draw[msg] (\xA,-12.45) -- node[lblbox,above]{Optional Peer Message} (\xM,-12.45);
\begin{pgfonlayer}{background}
  \node[annobox, fit={(\xA-0.1,-11.7) (\xM+0.0,-12.9)}, inner sep=2pt] (unsolbox) {};
\end{pgfonlayer}
\node[annolbl, anchor=north west] at ([xshift=3pt,yshift=0pt]unsolbox.north west) {Unsolicited Message};

% ==================== STEADY (repeat, fading) ====================
\draw[msg] (\xA,-14.15) -- node[lbl,above,font=\small\bfseries]{advance N} node[lbl,below]{grant next quantum} (\xM,-14.15);

\node[note, anchor=north east, font=\ttfamily\scriptsize] at (\xS-0.25,-14.25)
     {WHILE credits > 0:\\\quad credits -= 1;\\\quad EXEC TICK;};

\draw[msg] (\xM,-15.0) -- node[lbl,above]{ADVERTISE} node[lbl,below]{credit == 0} (\xA,-15.0);

\end{tikzpicture}}
  \caption{The steady-state co-emulation loop. The orchestrator grants the machine a bounded batch of tick credits; the machine reports back only once that batch drains or the firmware idles with none outstanding; the orchestrator then steps every peer in the interpreter, consults its own schedule, and grants the next quantum.}
  \label{fig:determinism}
\end{figure}
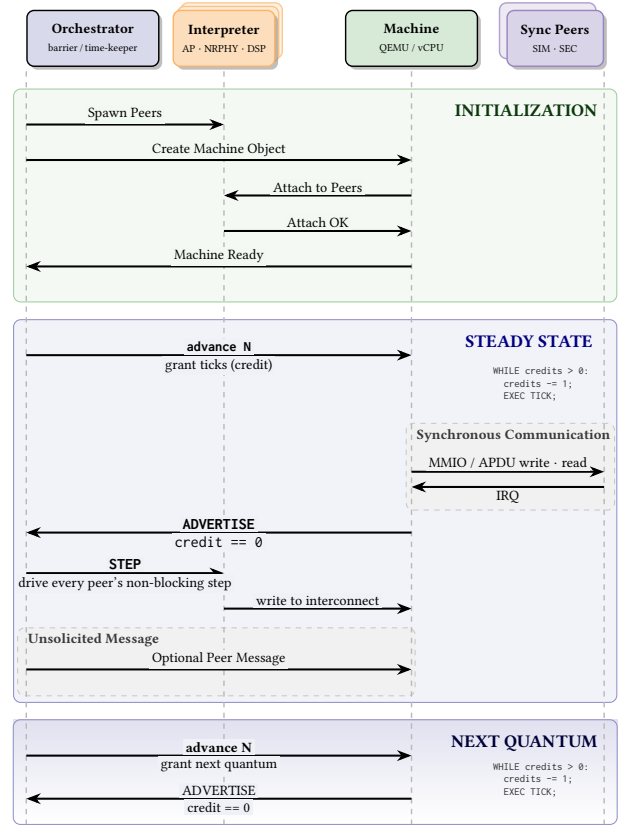

\subsection{Automating Model Construction}

Constructing a peer model decomposes into small, largely independent sub-problems.
For example, recovering the behavior of a single register, or the shape of a single message type, are local properties which make the construction itself eligible for automation.
We provide an agent the same introspection points we relied on manually: control of the testbed to start and stop the gNB and drive a PDU session; the ability to insert a patchable probe at a chosen program counter that dumps memory under a chosen condition; Ghidra access to the kernel, baseband, NRPHY, and NR DSP images; and the logs available from kernel modules, userspace applications, syscall traces, and kernel probes.
Given these primitives, the agent proposes a hypothesis about a peer's behavior, implements it as a model, and confirms or refutes it by A/B testing the model's output against the real device under identical stimulus, proceeding through a divide-and-conquer sequence of small objectives rather than a single monolithic reverse-engineering pass (\cref{fig:agent}).

\begin{figure}[t]
  \centering
  \resizebox{\linewidth}{!}{\input{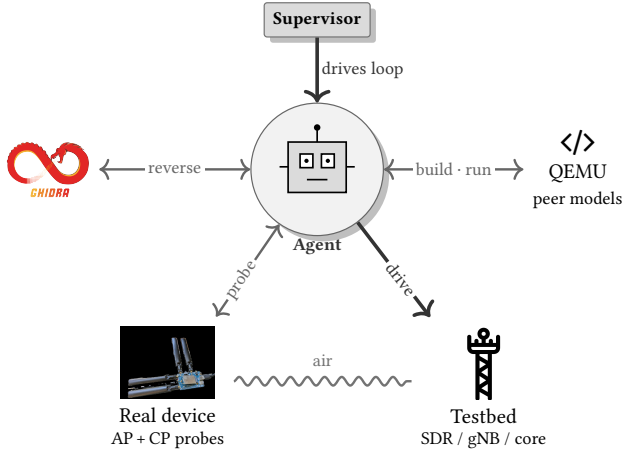}}
  \caption{Deriving the re-host from the real device. An agent reverses the firmware, probes the real device on a live testbed, and builds the QEMU emulation from the peer models it derives, modeling the device's real peripherals only.}
  \label{fig:agent}
\end{figure}

\subsection{Implementation}

We implement the machine in C and every peer in Python, embedded in the same process.
The guest and every peer therefore run on a single thread, not as separate loops each owning their own.
A per-tick call transfers control from the machine to the orchestrator and every peer in turn, so each can react before the tick continues.
Running everything on a single thread is a deliberate design choice.
It makes a non-issue of what would otherwise be two locks to coordinate: the Python interpreter's GIL and QEMU's big lock (BQL).
With only one thread, neither is ever contended; each is simply acquired to run a step and released immediately after.
The SIM is the one exception.
It is bridged through a direct blocking call rather than the per-tick step, since it cannot tolerate any delay in its reply.
Under \texttt{icount} this costs no virtual time, since the guest instruction count does not advance while the call blocks.
Peers with no firmware of their own, specifically the crypto engine, are instead modeled directly as synchronous MMIO.
The operation completes within the same write that rings its doorbell.

\section{Fidelity Demonstration}
\label{sec:evaluation}

This section empirically demonstrates that a peer- and interconnect-driven design re-hosts the UNISOC UDX710 baseband faithfully.
We pose two questions, one per experiment, and observe both through the same unified peer trace.
\begin{itemize}
  \item[] \textbf{Q1:} If both are driven from the same state, does the re-host take the same state transitions as real silicon?
  \item[] \textbf{Q2:} Can the re-host reach a full PDU session and carry genuine uplink and downlink IP packets?
\end{itemize}

These findings are empirical rather than exhaustive: for each transition reported below, we manually verified that the re-host reaches it at the same point in the procedure, and in the same way, as the real device.

\textbf{State-Machine Correlation.}
To answer \textbf{Q1}, we inject the same probe into the firmware itself, hooking its NAS state-change function directly; NAS is the control-plane protocol between the device and the core network, so its state machine is exactly the one registration and session setup drive.
Both environments therefore run the identical firmware binary and expose the identical state-change events, rather than two independent logging paths that happen to agree.
We start the real device and the re-host from the same fixed state and compare the state transitions each one reports after receiving SIB1 and camping on the cell.
Immediately after camping, both report the identical NAS 5GMM control-plane state transition, \texttt{[NGMM STATE CHANGE] MAIN [200:5->=1]!} followed by \texttt{[NGMM STATE CHANGE] DEREG [201:5->=1]!}, the same main-state and sub-state codes, at the same point in the procedure.
Later, once \texttt{RegistrationComplete} is sent, both report a second matching transition, \texttt{[NGMM STATE CHANGE] MAIN [200:1->=4]!} followed by \texttt{[NGMM STATE CHANGE] DEREG [201:5->=4]!}, again at the same point relative to authentication and security-mode setup.
We correlate these lines manually rather than automatically, but a match remains direct evidence that the two state machines take the same transition from the same state.

\textbf{Establishing Control- and Data-Plane Traffic.}
To answer \textbf{Q2}, a firmware log is not enough: reaching a PDU session and carrying IP traffic have to be shown crossing every peer boundary along the way, not reported by the firmware alone.

\textbf{Interconnect Dump Module.}
To get a boundary-level view, we embed an \textit{interconnect dump} module directly in each peer: every ingress and egress payload is written to a common capture file, and a Wireshark dissector decodes these dumps into readable frames.
Dumping the underlying DMA and ring-buffer I/O and indexing each transfer this way yields a single packet capture per run: SIM APDU exchanges, over-the-air control-plane messages, and AT commands, interleaved on one timeline in the order the orchestrator actually delivered them.
This lets us inspect the exact sequence of events across the SoC's peer boundaries in one place, rather than reconstructing it from independent logs, which is exactly the evidence this experiment needs.

We evaluate the interconnect dump along two axes, both visible in the same trace, since control-plane and data-plane traffic cross the identical NRPHY and AP boundaries: the control plane, the signaling that establishes a PDU session, and the data plane, the IP traffic that the session then carries.

\textbf{Control Plane.}
We compare the over-the-air RRC and NAS payloads the interconnect dump captured on the real device against those captured from the re-host for the same procedure, registration, authentication, and security-mode setup, and observe the identical payloads at each step.
The one exception is where cryptographic material is involved: the real SIM's sequence number for replay protection advances between runs, and the random challenge values it supplies are not fixed the way the re-host's are, so ciphered fields differ between the two runs; comparing the decrypted payload itself, however, still shows a match.

\textbf{Data Plane.}
We inject IP traffic in both uplink and downlink directions and track it across the same two boundaries.
For downlink, we inject a packet at the NRPHY ring; the baseband processes it through MAC, RLC, and PDCP, and delivers it to the AP, where it egresses through the kernel's virtual Ethernet interface.
For uplink, we inject the packet at the application-processor boundary instead and check for its egress at NRPHY, the reverse path.
Because the interconnect dump module records both boundaries on one timeline, the injected payload and its egress appear as directly comparable frames in the same PCAP capture.
A byte-identical match confirms the packet traversed the entire protocol stack.

Together, these demonstrations empirically show that the re-host reaches the same control-plane state as the real device for the 5G control-plane procedures we tested, and carries genuine, authentic control-plane and data-plane payloads across the full stack.
We provide a Wireshark trace collected from the emulator, covering both experiments, in \cref{sec:trace-appendix}.

\section{Discussion and Future Work}
\label{sec:discussion}

\textbf{Testing beyond 5G.}
Our re-host currently targets 5G NR only; LTE, 3G, and GSM depend on a different PHY co-processor we have not modeled to the same depth.
This is a testbed limitation more than a methodological one: reaching those generations requires the controlled radio testbed our requirements call for, and we have not yet deployed one that supports LTE.
The architecture already anticipates the extension: our LTE PHY peer currently answers only a band-sweep request with a truthful \enquote{no cell} response, enough to force the firmware's NR fallback path, but the same reverse-engineering and modeling approach we used for NRPHY applies directly once a suitable testbed is in place, and the same holds for the 3G and GSM co-processors.

\textbf{Generalizing Across UNISOC Basebands.}
We built and validated this re-host against the UDX710, but UNISOC reuses SoC components and layout across its baseband family, and we expect this effort to extend to other UNISOC devices.
We verified this empirically: downloading and comparing firmware images from other UNISOC basebands, e.g., the UMS9620, against the UDX710's own shows the same binary and module structure, the same NV and UBI partition layout, and the same driver code for peer-relevant components.
Peers bound to standardized, vendor-independent interfaces, the SIM's ISO-7816 controller is one, are reasonable candidates to carry over largely unchanged; peers bound to the radio stack, such as NRPHY, more plausibly need adjustment, though only to the degree the underlying co-processor generation actually differs.
Whether a new UNISOC baseband still requires the same amount of dynamic probing, or whether the models we have already recovered transfer with little or no adaptation, remains to be evaluated.

\textbf{Fuzzing at the Interconnect Boundary.}
Every peer channel is a well-defined interconnect boundary rather than an opaque region of guest memory, so we can inject arbitrary payloads at exactly the points an attacker would reach, over the air at NRPHY, or over the host interface at the AP.
Systematically fuzzing these boundaries, rather than manually driving a fixed reference session, is future work we intend to pursue.

\section{Related Work}
\label{sec:relwork}

Due to the security-critical nature of basebands, researchers have reverse engineered most baseband implementations and, in
some cases, developed on-device analysis for them: Shannon~\cite{golde2016breakingband,cama2018walk,berard2020basebanddebugger,grassi2021ota,silvanovich2023how,xing2023over,komaromy2024unburdened}, 
MediaTek~\cite{grassi2020mediatek,taszk2022csn1,komaromy2023basebanheimer}, and 
Qualcomm~\cite{burke2018hexagon,delugre2011qualcomm,gong2019qualcomm,gong2021hexagon}. 
For the UNISOC baseband, only one work by Makkaveev explores the modem image format and the NAS layer
parser~\cite{makkaveev2022unisoc}.

Building on the reverse engineering and on-device work,
the security community has started to explore how to re-host basebands for off-device testing with introspection.
BaseSafe~\cite{maier2020basesafe} uses selective emulation to only emulate specific functions of the MediaTek baseband.
Due to this over-approximation, only code in the baseband that does not depend on either state or peripherals can be emulated.
FirmWire~\cite{hernandez2022firmwire} is the first work on full-system baseband emulation for Shannon and MediaTek.
To boot the baseband, it models peripherals that are required for the boot. Then it systematically disables RTOS tasks which
crash due to missing hardware dependencies. 
This under-approximation leaves the baseband in a state in which protocol handlers can only be reached by directly interfacing 
with the RTOS queues.
BaseBridge~\cite{klischies2025basebridge} and 
FirmState~\cite{jeong2025firmstate} recover state from the real device to instantiate 
the emulated baseband in a specific state. While this allows them to 
instantiate the baseband with built-up state, the problem of peripheral handling 
needed to drive message sequences remains unsolved.
Loris~\cite{ranjbar2025stateful} limits emulation to a single control-plane task to explore state using
concolic execution. By design, it limits itself to a subset of the baseband's code with this over-approximation.
Our work is the first to tackle emulating the UNISOC baseband.
More importantly, it is the first to define the concept of peers,
needed to drive high-fidelity emulation of the baseband.

\section{Conclusion}
\label{sec:conclusion}

We present \unislop, the first systematic security analysis of a UNISOC baseband: gaining code execution, bypassing Quectel's firmware-integrity protection, and instrumenting the running firmware for tracing and introspection.
Building on this access, we introduce a peer- and interconnect-driven design for re-hosting the baseband as a deterministic emulation, reconstructing every SoC component the firmware depends on from real-device introspection rather than approximating it, and define fidelity operationally at the interconnect boundaries those peers communicate across.

We demonstrate this fidelity empirically: the re-host reaches the same control-plane state transitions as the real device from the same starting state, and carries genuine, authentic control-plane and data-plane traffic through a full PDU session, evidence we provide as a Wireshark trace alongside this paper.
The method already generalizes beyond a single target: static comparison across UNISOC firmware images shows the SoC components we modeled are shared across the baseband family, and the peer-driven design is what makes extending it to further protocol generations, further devices, and systematic fuzzing tractable rather than a rewrite.

\endgroup

\bibliographystyle{ACM-Reference-Format}
\bibliography{main}
\appendix

\section{Unlocking ADB}
\label{sec:unlock-appendix}

We download the latest firmware available for the RM500U-CNV through Quectel's support channels.
The firmware images are not publicly hosted; in practice, access requires requesting firmware through Quectel's forum or support moderators.

After receiving the firmware image, we find that it is not encrypted and can be unpacked directly.
Initial inspection shows that the platform contains a Linux-based ARM64 subsystem:

\begin{center}
\begin{minipage}{0.9\linewidth}
\begin{lstlisting}
Linux udx710-module 4.14.98 #1 SMP PREEMPT Tue Jul 1 14:17:41 UTC 2025 aarch64 GNU/Linux
\end{lstlisting}
\end{minipage}
\end{center}

The unpacked filesystem contains several binaries relevant for further analysis. 
Specifically, we identify components responsible for AT command handling (\texttt{atrouter}), modem control and device configuration (\texttt{modem\_control}), and Android Debug Bridge support (\texttt{adbd}).
The presence of \texttt{adbd} suggests a path to interactive shell access, however, no ADB interface is exposed by default.

To understand why ADB is present but not reachable, we search the firmware for ADB-related references.
This leads us to the proprietary \texttt{atrouter} binary, which implements the AT command \texttt{AT+QADBKEY}.
Reverse engineering this command shows that the firmware derives an ADB unlock key from the device serial number and a salt embedded in the binary.

For firmware version \texttt{5G\_MODEM\_21A\_W23.38.3\_P28}, the procedure starts by querying the modem for the ADB challenge over the AT interface exposed on \texttt{/dev/ttyUSB4}:

\begin{center}
\begin{minipage}{0.9\linewidth}
\begin{lstlisting}
AT+QADBKEY?
ABCDEF_7_13_27
\end{lstlisting}
\end{minipage}
\end{center}

We then compute the corresponding unlock key using the salt recovered from the \texttt{atrouter} binary:

\begin{center}
\begin{minipage}{0.9\linewidth}
\begin{lstlisting}
/usr/bin/openssl passwd -1 -salt "QUE_V002" "ABCDEF_7_13_27"
$1$QUE_V002$rjihvi4Ai1D3AswNWdCt7.
\end{lstlisting}
\end{minipage}
\end{center}

We submit the computed key and reconfigure the USB composition to expose ADB:

\begin{center}
\begin{minipage}{0.9\linewidth}
\begin{lstlisting}
AT+QADBKEY="rjihvi4Ai1D3AswNWdCt7"
AT+QCFG="usbcfg",0x2C7C,0x0800,1,1,1,1,1,1,0
\end{lstlisting}
\end{minipage}
\end{center}

After USB reconfiguration, ADB becomes accessible and gives a root shell on the Linux subsystem.

\section{Execution Tracing Instrumentation}
\label{sec:tracing-appendix}

This appendix gives the concrete form of the execution-tracing instrumentation summarized in \cref{sec:observing}.
Each rewritten call site branches to a trampoline (\cref{fig:trampoline}) that saves the caller's context, invokes the instrumentation target (\cref{fig:cov-hit}), restores the context, and only then branches to the original target.
The target stays inert until the modem reports boot completion, after which it re-executes the overwritten instructions and folds the return address into an index into the 65\,kB trace map.

% [H]: these two listings are the only content of the appendix page; letting
% them float would leave a float-only page, which drops the surrounding text.
\begin{figure}[H]
\centering
\begin{minipage}{0.95\linewidth}
\begin{lstlisting}[numbers=none]
_start:
stmdb sp!, {r0, r1, r2, r3, r4, r5, r6, r7, r8, r9, sl, fp, lr}
bl 0 <instrumentation_target>
ldmia.w sp!, {r0, r1, r2, r3, r4, r5, r6, r7, r8, r9, sl, fp}
bl e00000 <original_target>
\end{lstlisting}
\end{minipage}
\caption{Trampoline emitted at each instrumented fixed-target call site. It saves the execution context, calls the instrumentation target, restores the context, and branches to the original call target.}
\label{fig:trampoline}
\end{figure}

\begin{figure}[H]
\centering
\begin{minipage}{0.95\linewidth}
\begin{lstlisting}[language=C]
__attribute__((naked))
void _start(void) {
  save_lr();

  int lr_addr = 0;

  /* Load context from reserved memory region */
  cov_ctx_t ctx = {
   .hdr = (cov_hdr_t *)COV_HDR_ADDR,
   .map = (char *)COV_MAP_ADDR,
  };

  /* Tracing is enabled after boot completes */
  if (!is_cov_ready(&ctx)) {
      goto retfn;
  }

  /* Re-execute overwritten instructions */
  PATCHED_INSN(lr_addr);
  /* Record tracing instrumentation hit */
  cov_hit(&ctx, lr_addr);

retfn:
  ret();
}

/* Tracing hit instrumentation */
void cov_hit(cov_ctx_t *c, int lr) {
  /* Auxiliary helpers */
  c->hdr->iter_count += 1;
  c->hdr->last_lr = lr;

  /* Reconstructible IDX mapping */
  unsigned int idx = map(lr) % (COV_MAP_SIZE - 1u);

  c->map[idx]++;
}
\end{lstlisting}
\end{minipage}
\caption{Instrumentation target invoked by the trampoline. It records a trace hit only once the modem reports boot completion, re-executes the instructions overwritten by the patch, and folds the return address into the trace map.}
\label{fig:cov-hit}
\end{figure}

\section{Peer Trace Comparison}
\label{sec:trace-appendix}

\cref{tab:rrc-trace} lists the NR RRC messages from the Wireshark trace referenced in \cref{sec:evaluation}, filtered to the \texttt{nr-rrc} protocol layer.
The sequence covers cell camping, RRC connection setup, NAS registration and authentication, security-mode configuration, and PDU session establishment, in the same order and at the same points as on the real device.

\begin{table*}[]
  \centering
  \footnotesize
  \caption{NR RRC messages in the co-simulation Wireshark peer trace, filtered to the \texttt{nr-rrc} protocol layer with \texttt{tshark}.}
  \label{tab:rrc-trace}
  \begin{tabular}{@{}r r l l l p{8.4cm}@{}}
    \toprule
    \textbf{No.} & \textbf{Time (s)} & \textbf{Src} & \textbf{Dst} & \textbf{Protocol} & \textbf{Info} \\
    \midrule
    661 & 38.877 & nrphy & cp & NR RRC & SIB1 \\
    683 & 38.957 & cp & nrphy & NR RRC & RRC Setup Request \\
    685 & 38.958 & nrphy & cp & NR RRC & RRC Setup \\
    709 & 39.034 & cp & nrphy & NR RRC/NAS-5GS & RRC Setup Complete, Registration request \\
    722 & 42.516 & nrphy & cp & NR RRC/NAS-5GS & DL Information Transfer, Authentication request [51-bytes] \\
    739 & 42.718 & cp & nrphy & NR RRC/NAS-5GS & UL Information Transfer, Authentication response \\
    741 & 42.870 & nrphy & cp & NR RRC/NAS-5GS & DL Information Transfer, Security mode command [24-bytes] \\
    752 & 42.905 & cp & nrphy & NR RRC/NAS-5GS/NAS-5GS & UL Information Transfer, Security mode complete, Registration request \\
    754 & 43.212 & nrphy & cp & NR RRC & Security Mode Command [9-bytes] \\
    765 & 43.244 & cp & nrphy & NR RRC & Security Mode Complete \\
    771 & 44.814 & nrphy & cp & NR RRC/NAS-5GS & DL Information Transfer, Registration accept [60-bytes] \\
    833 & 44.870 & cp & nrphy & NR RRC/NAS-5GS & UL Information Transfer, Registration complete \\
    844 & 45.171 & nrphy & cp & NR RRC/NAS-5GS & DL Information Transfer, Configuration update command [20-bytes] \\
    908 & 45.219 & cp & nrphy & NR RRC/NAS-5GS & UL Information Transfer, Configuration update complete \\
    970 & 45.271 & cp & nrphy & NR RRC/NAS-5GS & UL Information Transfer, UL NAS transport, PDU session establishment request \\
    988 & 66.836 & nrphy & cp & NR RRC/NAS-5GS & DL Information Transfer, DL NAS transport, PDU session establishment accept [87-bytes] \\
    1053 & 66.886 & cp & nrphy & NR RRC/NAS-5GS & UL Information Transfer, UL NAS transport, PDU session establishment request \\
    1062 & 74.534 & nrphy & cp & NR RRC/NAS-5GS & RRC Reconfiguration, DL NAS transport, PDU session establishment accept [389-bytes] \\
    1135 & 74.601 & cp & nrphy & NR RRC & RRC Reconfiguration Complete \\
    1201 & 79.871 & cp & nrphy & NR RRC/NAS-5GS & UL Information Transfer, Service request \\
    \bottomrule
  \end{tabular}
\end{table*}

\end{document}